\documentclass[11pt]{iopart} 
\usepackage{graphicx}
\usepackage{cite}
\usepackage{amsfonts}
\usepackage{array}

\def\rvec{{\vec{r}}}
\def\NN{\mathbb{N}}
\begin{document}

\title[Local Persistence in the Directed Percolation Universality Class]
      {Local Persistence in the Directed Percolation Universality Class}
\author{Johannes Fuchs$^1$, J{\"o}rg Schelter$^1$, Francesco Ginelli$^2$,\\ and Haye Hinrichsen$^1$}
\address{$^1$ Universit\"at W\"urzburg, Fakult\"at f\"ur Physik und Astronomie, 
         \\ D-97074 W\"urzburg, Germany}
\address{$^2$ Institut des Syst\`emes Complexes 57/59, rue Lhomond F-75005 Paris, France \\ and CEA/Saclay,
         F-91191 Gif-Sur-Yvette, France}

\begin{abstract}
We revisit the problem of local persistence in directed percolation, reporting improved estimates of the persistence exponent in 1+1 dimensions, discovering strong corrections to scaling in higher dimensions, and investigating the mean field limit. Moreover, we examine a graded persistence probability that a site does not flip more than $m$ times and demonstrate how local persistence can be studied in seed simulations. Finally, the problem of spatial (as opposed to temporal) persistence is investigated.
\end{abstract}

\submitto{Journal of Statistical Mechanics: Theory and Experiment}

\maketitle
\def\pl{{P_\ell}}
\def\tl{{\theta_\ell}}

\parskip 1mm 
%
\section{Introduction}
\label{intro}
%
Switching on the coffee machine, how long does it take until you smell the coffee for the first time? In statistical physics questions of this type are known as first-passage problems~\cite{Redner}. Equivalently one may ask for the \textit{persistence probability} that someone, after turning on the coffee machine, does not smell the coffee for a certain period of time. In more general terms, the \textit{persistence probability} is defined as the probability that a certain observable in a random process does not cross its expection value from the initial state until time $t$. 

In the last decade persistence has been studied both theoretically and experimentally in a large variety of systems~[2-75].
There are two main categories of persistence studies. For global observables such as the total magnetization or the average particle density, one studies the \textit{global persistence probability} $P_g(t)$ that an order parameter does not cross its expectation value up to time~$t$~\cite{Majumdar96b,Stauffer96,Lee97,Menyhard97,Schuelke97,Zanette97a,Dornic98,Drouffe98,Oerding98,Majumdar03,Saharay03,Silva03,Sen04,Silva05,Fernandes06a,Fernandes06b}. In critical dynamical systems it is observed that the global persistence probability decays algebraically as $P_g(t)\sim t^{-\theta_g}$. For Markovian dynamics one can show that the global persistence exponent can be expressed by a scaling relation $\theta_g z = \lambda-d+1+\eta/2$, where $d$ is the spatial dimension while $\eta$ and $\lambda$ are the static and the autocorrelation exponents, respectively. 

The other category of persistence studies, on which we will focus in the present work, deals with \textit{local} observables such as the occupation number or spin orientation at individual sites. Here, one is interested in the \textit{local persistence probability} $\pl(\vec{r},t)$ that the local state at a given position $\vec{r}$ in space has not changed up to time $t$. For example, in spin models $\pl(\vec{r},t)$ may be defined as the probability that a spin at a given site does not flip until time $t$. If the system is translationally invariant, the persistence probability does not depend on $\vec{r}$ so that $\pl(t)$ is just the fraction of spins that did not flip until time $t$. 

In many systems evolving towards a scale-free state the local persistence probability is found to decay algebraically as
\begin{equation}
\label{PowerLaw}
\pl(t) \sim t^{-\tl}\,,
\end{equation} 
where $\tl$ is the so-called local persistence exponent~\cite{Bray94,Derrida94,Stauffer94,Cardy95,Derrida95,Derrida96a,Derrida96b,Derrida97,Hennecke97,Stauffer97,Hinrichsen98a,Hinrichsen98b,Manoj03,Menon03,Saharay03,Das04,Sen04,Dutta05}. This exponent is generally different from $\theta_g$ and seems to have certain universal features, i.e., its numerical value may coincide in various models with continuous phase transitions belonging to the same universality class. However, in contrast to $\theta_g$, which can be expressed in terms of the bulk exponents by a scaling relation, $\tl$ seems to be an independent exponent. So far there are only very few exact results~\cite{Derrida94,Cardy95,Derrida95,Derrida96a,Derrida96b,Krapivsky98,Howard98}. One of the most pathbreaking results was obtained by Derrida {\it et al.}~\cite{Derrida95,Derrida96a}, who were able to compute the persistence exponent~$\tl$ in the one-dimensional Ising and Potts model with heat bath dynamics. This exact solution maps the Glauber-Ising dynamics to a time-reversed coagulation process with a special boundary condition and relates $\tl$ to a universal amplitude close to the boundary. 

Persistence was also studied in various other contexts, including random walks and Gaussian processes~\cite{Majumdar96a,Majumdar96c,Bauer99,Ehrhardt04}, fluctuating interfaces and surface growth~\cite{Krug97,Kallabis99,Constantin03,Constantin04a,Constantin04b,Majumdar05,Singha05}, systems with shear flow~\cite{Bray04,Bray05,Rapapa06}, models with algebraic long-range interactions~\cite{Zanette97b,Ispolatov99}, and disordered systems~\cite{Newman99,Silva04,Paul05,Pleimling05}.  Moreover, various authors have improved and generalized the concept of persistence in different ways~\cite{Cueille97,Majumdar98a,Majumdar98b,Cueille99,Baldassarri99,Sire00,Krishnamurthy03,Merikoski03,Lima04,Gonos05,Kumar05,Shukla05,Baldassarri99}. Some predictions were confirmed experimentally and successfully applied to financial data~\cite{Yurke97,Zheng02,Merikoski03,Ren04,Constantin05,Dougherty05,Silva06}.

\begin{figure}
\begin{center}
\includegraphics[width=110mm] {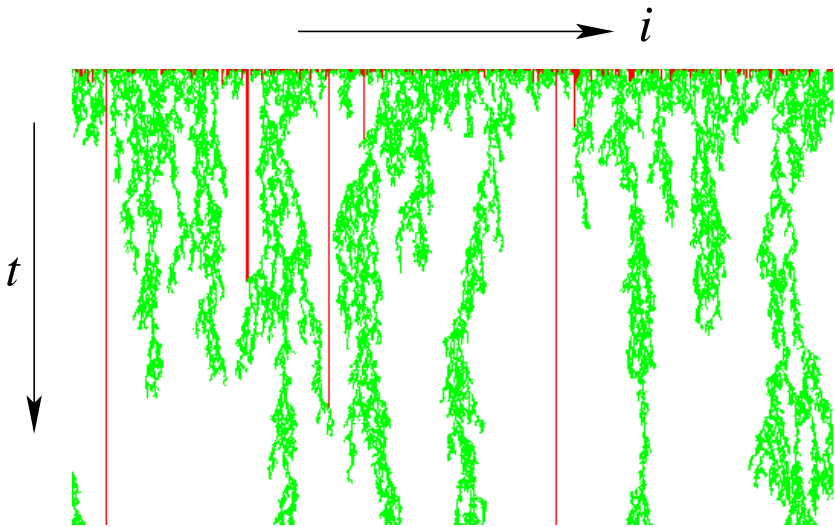}
\vspace{-2mm}
\end{center}
\caption{\small Persistent sites (red) in a 1+1-dimensional critical DP process (green). }
\label{fig:demo}
\end{figure}

Within this broad range of models and applications, the present paper addresses a particular issue, namely, the problem of local persistence in the directed percolation universality class (DP)~\cite{Jensenn81, Kinzel85, Hinrichsen00}, the simplest and paradigmatic class of non-equilibrium phase transitions from a fluctuating phase into one or several absorbing states. In models with absorbing states local persistence may be defined as the probability that a given site -- monitored shortly after the initial state up to time~$t$ -- remains locally absorbing (inactive). For example, in a $d$+1-dimensional DP process starting with a fully occupied infinite lattice, $\pl(t)$ would be the density of sites that have never been activated up to time $t$, excluding the initial state at $t=0$. A typical run of a 1+1-dimensional DP process is shown in Fig.~\ref{fig:demo}, where locally persistent sites are represented as vertical red lines.

In the context of absorbing phase transitions local persistence was studied for the first time in Ref.~\cite{Hinrichsen98a}, where the fraction of unaffected sites in a DP process was investigated numerically. At criticality in 1+1 dimensions the persistence probability was found to decay algebraically as $\pl(t) \sim t^{-\tl}$ with an exponent $\tl^{1D}=1.50(2)$ which is apparently independent of the bulk exponents $\beta,\nu_\parallel,\nu_\perp$. Later Albano and Mu{\~n}oz~\cite{Albano01} extended these studies to higher dimensions. Investigating the ZGB model at its DP transition point, they reported the estimates $\tl=1.50(1)$ in one \textit{and} two, $\tl=1.33(3)$ in three, and $\tl=1.15(5)$ in four spatial dimensions. \\

\noindent
In the present work we revisit the problem of local persistence in DP from various perspectives. The main results are:
\begin{itemize}
\item[a)] The algebraic decay in Eq.~(\ref{PowerLaw}) is confirmed within a mean field approximation. However, in mean field the mechanism leading to a power-law behavior is shown to differ from the one in low-dimensional systems.
\item[b)] We provide a more accurate estimate of $\tl$ in 1+1 dimension. In higher dimensions, however, it is shown that long transients make a reliable numerical estimation impossible, questioning previous estimates reported in Ref.~\cite{Albano01}.
\item[c)] We discuss the scaling properties of local persistence, suggesting a formulation in the continuum limit and investigating the concept of graded persistence.
\item[d)] For the first time local persistence is investigated in seed simulations, where a different local persistence exponent is observed.
\item[e)] We analyse spatial (as opposed to temporal) persistence and find a relation between the corresponding exponents.
\end{itemize}

\section{Mean field calculation}
%
In directed percolation, the mean field approximation is expected to be valid above the upper critical dimension $d_c=4$. However, the results of this section indicate that, as in other reaction-diffusion models \cite{Cardy95}, the local persistence exponent for $d_c<d<\infty$ turns out to be $d$-dependent and non-universal. To illustrate this, we shall write down mean field approximations for two different lattice models in the DP universality class: the contact process \cite{Liggett85} and bond directed percolation \cite{Adler83}. 

\subsection{Mean field approximation for the contact process}
%
The contact process is a lattice model with random-sequential dynamics, where particles create offspring ($A \to 2A$) at rate $\lambda$ and self-annihilate ($A \to \emptyset$) at rate $1$. The mean-field evolution equation of this process reads
\begin{equation}
\partial_t \rho = (\lambda - 1) \rho(t) - \lambda \rho^2(t) \,,
\label{MF1}
\end{equation}
where $\rho(t)$ is the coarse-grained density of active sites. In addition, let $\psi(t)$ be the density of persistent sites that have not been activated up to time $t$. Since the mean-field limit ignores fluctuations, the loss of persistent sites is proportional to the frequency of new activations multiplied by the concentration of persistent sites, leading to the additional differential equation
\begin{equation}
\partial_t \psi(t) = - \lambda \rho(t) \psi(t)\,.
\label{MF2}
\end{equation}
Setting $\rho(0)=\psi(0)=1$ the solution at the critical point $\lambda = \lambda_c = 1$ is given by
\begin{equation}
\rho(t) = \psi(t) = \frac{1}{1+t} \sim t^{-1}\,.
\label{MFeq}
\end{equation}
This means that for the contact process the persistence exponent within mean field theory is given by
\begin{equation}
\tl^{\rm MF} = 1
\label{MFexp}
\end{equation} 
in any space dimension $d \geq d_c$. It is also instructive to study the mean field solution above and below the critical point. Here the solution reads
\begin{equation}
\rho(t) = \psi(t) = \frac{\lambda-1}{\lambda-\exp[(1-\lambda)t]}\,.
\end{equation}
In the inactive phase $\lambda < 1$, the density of active sites decays exponentially while the density of persistently inactive sites saturates for large times at a constant:
\begin{eqnarray}
\rho(t) &\simeq& (1-\lambda) \exp[-(1-\lambda) t]\,,\\
\psi(t) &\to& 1-\lambda\,.
\end{eqnarray}
In the active phase $\lambda > 1$, however, the roles between $\rho$ and $\psi$ are exchanged 
and one obtains
\begin{eqnarray}
\rho(t) &\to& 	 \frac{\lambda-1}{\lambda}\,, \\
\psi(t) &\simeq& \frac{\lambda-1}{\lambda}\,\exp[-(\lambda-1) t] \,.
\end{eqnarray}
Since the time scale in the exponential functions is expected to scale as $|\lambda-1|^{-\nu_\parallel}$ and the density of persistent sites should saturate in the second case at a level proportional to $|\lambda-1|^{-\tl\nu_\parallel}$, these results are compatible with the well-known mean field exponents $\beta^{\rm MF}=\nu_\parallel^{\rm MF}=1$ of directed percolation.\\

\subsection{Mean field approximation for bond directed percolation}

In order to illustrate in detail that the persistence exponent in finite dimensions above $d_c$ is generally a $d$-dependent non-universal exponent, let us now consider the example of directed bond percolation. In bond DP on a tilted square lattice each lattice site is connected to $2d$ neighboring sites via directed bonds which are open with probability $p$ and closed otherwise. Assuming the states of these neighboring sites to be uncorrelated, the probability that the updated site becomes active is $1-(1-\rho)^k$, where $k \in \{0,\ldots 2d\}$ is the number of open bonds. Since the probability to find $k$ out of $2d$ bonds opened is given by ${{2d} \choose k} p^k (1-p)^{2d-k}$, the average density of active sites after the update, $\rho_{\rm new}$, within mean field approximation reads
\begin{eqnarray}
\rho_{\rm new} &=& \sum_{k=0}^{2d} \,  {{2d} \choose k} p^k (1-p)^{2d-k} \, [1-(1-\rho)^k] \nonumber \\
&=& 1-(1-p\rho)^{2d}
\;=\; 2dp\rho - (2d^2-d)p^2\rho^2 + O(\rho^3)\,.
\end{eqnarray}
At the mean-field critical point $p_c =1/2d$ this expression expanded to second order in the continuum limit reduces to the differential equation $\dot\rho = -\frac12(1-\frac{1}{2d}) \rho^2$ with the asymptotic solution $\rho(t) \sim \frac{2}{t(1-1/2d)+2}$. Given this solution at criticality the density of persistent sites , to leading order in $\rho$, evolves according to the differential equation
\begin{equation}
\partial_t \psi(t) = - \frac12 \rho(t) \psi(t)\,,
\label{MF2bis}
\end{equation}
where the factor $\frac12$ stems from the fact that in bond-DP a given site is updated only every second time step. The solution is 
\begin{equation}
\psi(t)\;=\;\left[\frac{2 d-1}{4d}\, t + 1 \right]^{\frac{2 d}{1-2 d}} 
\;\sim\; t^{\frac{2 d}{1-2 d}} \,, 
\end{equation}
hence the local persistence exponent is given by
\begin{equation}
\tl = \frac{2d}{2d-1} \stackrel{d \rightarrow \infty}{\to} 1\,.
\label{MFexp2}
\end{equation}
This result differs from the one for the contact process in Eq.~(\ref{MFexp}). It demonstrates that above the upper critical dimension, where the mean field approximation is expected to give correct answers, $\tl$ is a non-universal model-dependent quantity which generally depends on the dimensionality. Similar non-universal results were obtained for other reaction-diffusion models such as pair annihilation $2A\to\emptyset$, see e.g.~\cite{Cardy95}.

We finally note that the simple mean field results (\ref{MFexp}), (\ref{MFexp2}) rely on the assumption that the local density of active sites is spatially homogeneous, i.e., $\rho(\vec{x},t)$ takes the same value $\rho(t)$ both close and far away from persistent sites. In the full model this assumption is expected to be valid only in the limit $d\to\infty$. In finite dimensions above $d_c$, however, the density of active sites next to persistent ones is expected to be lower because a persistent site itself can be regarded as a defect of vanishing activity. These (anti)correlations may effectively change the density $\bar{\rho}(t)$ of active sites near the persistent ones, i.e. $\bar{\rho}(t)= \gamma_d \rho(t)$ with $\gamma_d \to 1$ for $d \to \infty$. Correspondingly, one should include this factor in Eqs. (\ref{MF2}) and (\ref{MF2bis}), leading to a further correction to the mean-field exponent $\tl \to \gamma_d \tl$.

The non-universality of $\tl$ for $d>d_c$ shows that the local persistence exponent differs qualitatively from the other bulk exponents of DP which are constant above the upper critical dimension. Apparently the Gaussian fixed point, which governs the critical behavior above $d_c$, does not impose constraints on the value of $\tl$. 

\section{Numerical estimation of $\tl$}

In Ref.~\cite{Hinrichsen98a} the local persistence exponent of 1+1-dimensional DP was determined by simulating a 1+1-dimensional Domany-Kinzel cellular automaton~\cite{DKModel,Kinzel85}, leading to the estimate $\tl=1.50(2)$. This value suggested that $\tl$ is probably unrelated to the other bulk exponents of DP, $\beta,\nu_\parallel$ and $\nu_\perp$. Similar estimates were found in other models for DP~\cite{Albano01}, suggesting that $\tl$ is universal in 1+1 dimensions. Moreover, it was speculated whether $\tl$ is exactly equal to the rational value $3/2$ or not.

\begin{figure}
\begin{center}
\includegraphics[width=100mm] {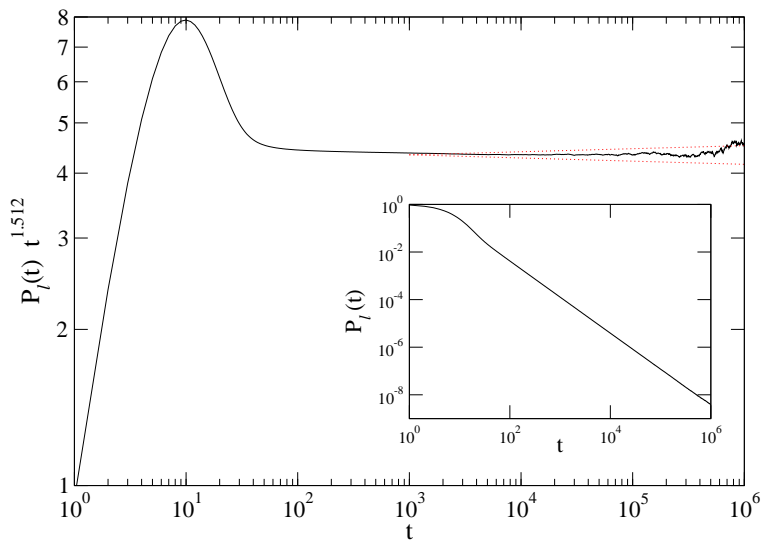}
\vspace{-3mm}
\end{center}
\caption{\small High-precision measurement of local persistence in the 1d contact process. The plot shows $\pl(t)$ multiplied by $t^\tl$ with $\tl=1.512$. The red dotted lines indicate the error margin of the estimated exponent $\pm 0.006$. The inset shows the raw data of the simulation.}
\label{fig:decay1d}
\end{figure} 

To address this question we performed extensive Monte-Carlo simulations at the critical point of the 1+1-dimensional contact process~\cite{Liggett85}, ranging up to $t=10^6$ time steps and averaged over $2.4 \times 10^7$ independent runs. This enormous number of sweeps is necessary to obtain reliable averages for the persistence probability which decreases down to $10^{-8}$. As shown in Fig.~\ref{fig:decay1d} the results confirm the anticipated power-law decay of $\pl(t)$. Moreover, the local persistence exponent is estimated by
\begin{equation}
\tl = 1.512(6)
\end{equation} 
so that we can now safely exclude the scenario of a rational value $3/2$. A similar numerical result was obtained for directed bond percolation with parallel updates. 

\begin{figure}
\centering\includegraphics[width=110mm] {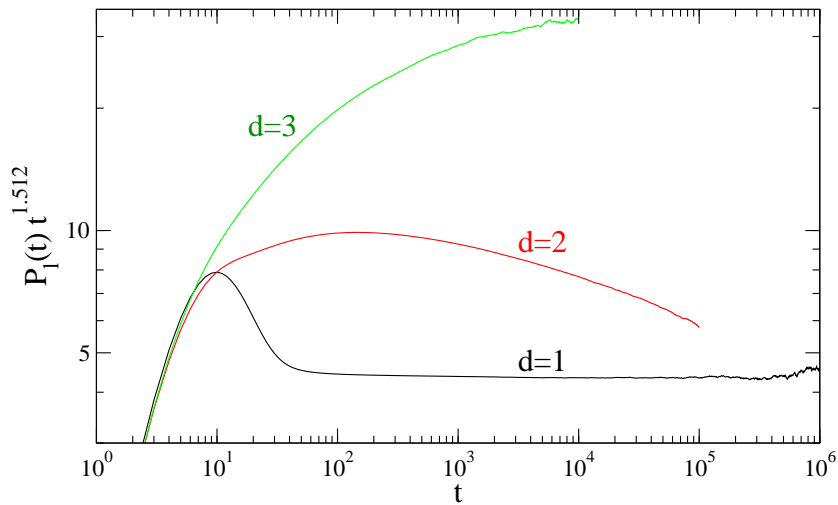}
\vspace{-3mm}
\caption{\small Local persistence probability multiplied by $t^{1.512}$ in the critical contact process in 1, 2, and 3 spatial dimensions.}
\label{fig:cpdim}
\end{figure} 

\begin{table}[b]
\begin{center}
\begin{tabular}{|c||c|c|c|c|c|}\hline
model & bond-DP	& ZGB 		& CP  		& bond-DP        \\ \hline 
ref.  & \cite{Hinrichsen98a} & \cite{Albano01}  & this work    & this work   \\ \hline 
$d=1$ & 1.50(2) &  1.50(1) 	&   1.512(6)	&  1.517(10)     \\ 
$d=2$ & --  	&  1.50(1)	&  $>1.62$      &  $>1.58$       \\ 
$d=3$ & --	&  1.33(3) 	&  $>1.5$ 	&  $>1.4$        \\ 
$d=4$ & --  	&  1.15(5)	&  --		&  $>1.1$        \\ \hline
\end{tabular} 
\end{center}
\label{Table:Exponents}
\caption{\small Estimates for $\tl$ in directed percolation. The inequalities for higher dimensions rely on the assumption that the curvature of the data shown in Fig.~\ref{fig:cpdim} does not change sign as $t \to \infty$.}
\end{table}

The measurement of persistence in higher dimensions is extremely difficult since finite-size effects become increasingly relevant. For example, in three dimensions the simulation time of $10^6$ time steps would require about $10^9$ lattice sites in order to exclude finite-size effects. In addition, a large number of runs is needed to obtain reasonable statistical averages. Fig.~\ref{fig:cpdim} shows $\pl(t)$ measured in the one-, two- and three-dimensional contact process at criticality within a temporal window where finite-size effects can be safely excluded. As can be seen, the lines for $d=2,3$ bend downward. Therefore, a reliable estimation of the critical exponents is impossible; judging from the numerics it is not even clear whether $\pl(t)$ exhibits an asymptotic power law at all. However, assuming that the asymptotic decay \textit{is} algebraic and that the curvature does not change sign as $t \to \infty$ the simulations allow us to specify lower bounds for $\tl$, see Table~\ref{Table:Exponents}. Note that the lower bounds for two spatial dimensions are actually larger than the $d=1$ estimate, while they become smaller for $d=3,4$, a somehow odd result that may indeed indicate abnormally long transient effects.   

To summarize, we obtain a reliable estimate for $\tl$ in 1+1 dimensions while we are unable to confirm or disprove a possible asymptotic power-law decay in higher dimensions. However, the results rule out a conjectured numerical coincidence of $\tl$ in 1+1 and 2+1 dimensions (called \textit{superuniversality}) as previously speculated in Ref.~\cite{Albano01}.

\section{Scaling properties}
%
The observed power-law behavior in 1+1 dimensions suggests to describe the interplay of several parameters by phenomenological scaling laws. In this section we discuss such scaling laws on different levels of complexity.

\subsection{Off-critical and finite-size scaling}

\begin{figure}
\begin{flushright}
\includegraphics[width=150mm]{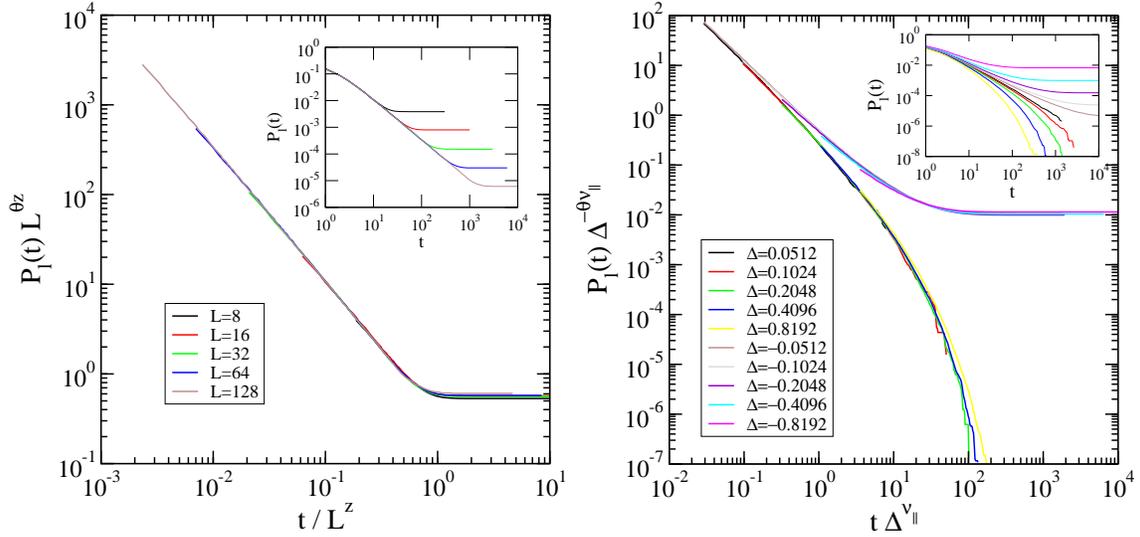}\end{flushright} \vspace{-5mm}
\caption{\small Local persistence probability in finite-size (left) and off-critical simulations (right) in the 1+1-dimensional contact process. The graphs demonstrate data collapses according to the scaling form~(\ref{GeneralScalingForm}), while the insets show the original data. }
\label{fig:fsoff}
\end{figure} 

The simplest extension is to include the lateral system size $L$ and the distance from criticality $\Delta=\lambda-\lambda_c$ as additional parameters. In analogy to other quantities exhibiting DP scaling one expects an asymptotic scaling law of the form
\begin{equation}
\label{GeneralScalingForm}
\pl(t) \;\simeq\; t^{-\tl} \, F\Bigl(\frac{t}{L^z},\, t\Delta^{\nu_\parallel}\Bigr)\,,
\end{equation}
where $F$ is a scaling function and $z=\nu_\parallel / \nu_\perp \simeq 1.581$ is the dynamical exponent of DP in 1+1 dimensions. In order to demonstrate the validity of this scaling form numerically, we consider two special cases:
\begin{itemize}
\item[(a)]
\textit{Finite-size scaling:} A finite 1+1-dimensional contact process at criticality $\Delta=0$ reaches the absorbing state within finite time so that there is a finite probability for persistent sites to survive forever. Therefore, the persistence probability saturates at a constant value. Plotting $\pl(t)L^{\tl z}$ against $t/L^z$ all curves collapse onto a single one, as demonstrated in Fig.~\ref{fig:fsoff}.
\item[(b)]
\textit{Off-critical simulations:} For an infinite 1+1-dimensional contact process below (above) the percolation threshold, we expect the persistence probability to saturate (decay exponentially). According to the scaling form~(\ref{GeneralScalingForm}) the curves should collapse in both cases when $\pl(t) \Delta^{-\tl \nu_\parallel}$ is plotted against $t\Delta^{\nu_\parallel}$. As shown in the right panel of Fig.~\ref{fig:fsoff}, this is indeed the case.
\end{itemize}
The successful data collapses confirm conventional power-law scaling in 1+1 dimensions. Whether similar scaling laws apply in higher dimensions is still an open question.

\subsection{Local persistence as a quantity depending on two time parameters}

In models with absorbing states local persistence is usually introduced as the probability that a given site is not activated until time $t$, excluding the initial configuration at $t=0$ where all sites are active. This definition implicitly suggests that the persistence probability $\pl(t)$ depends on only a \textit{single} time parameter~$t$. However, with only one parameter $t$ it is impossible to define a meaningful continuum limit of the persistence probability. For example, a fully occupied initial state in the lattice model translates into an infinite $\delta$-peak-like density of active sites at $t=0$ in the continuum description so that the probability for a given site to remain inactive in the interval $0 < \tau \leq t$ would be zero. Therefore, the only way to introduce persistence consistently is to define $\pl$ as the probability that a given site remains inactive in some interval between time $t_0>0$ and time $t$. This means that local persistence should be defined as a quantity $\pl(t,t_0)$ that depends on \textit{two} time parameters $t_0$ and $t$ which reduces to the previous definition by setting $t_0:=1$.

\begin{figure}
\begin{flushright}
\includegraphics[width=145mm]{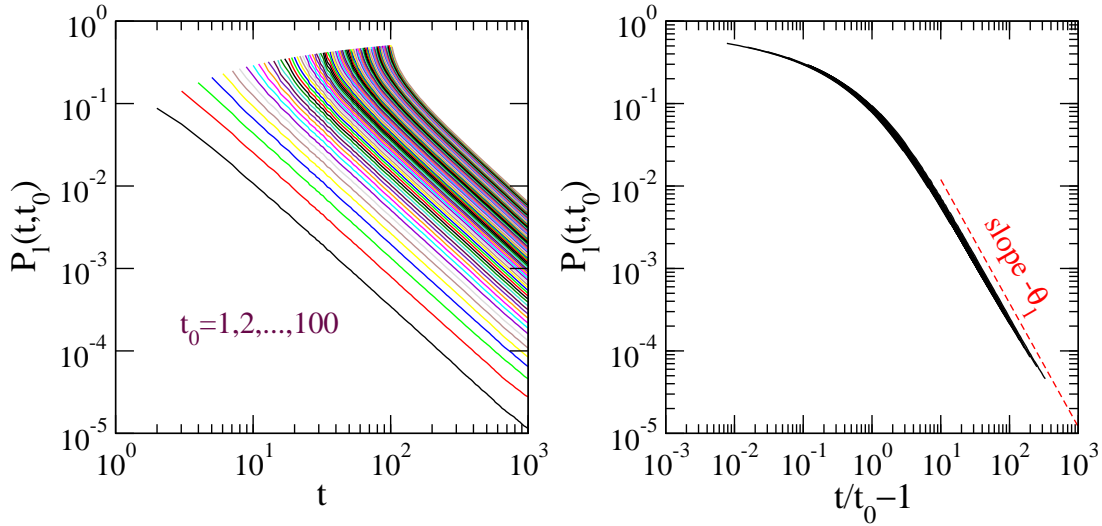}\end{flushright}
\caption{\small Measurement of the persistence probability $\pl(t,t_0)$ in a 1+1-dimensional critical contact process as a function of two variables $t$ and $t_0$. The left panel shows the original data while the right one demonstrates a data collapse according to Eq.~(\ref{CriticalScalingForm}).}
\label{fig:twotimes}
\end{figure} 

The left panel of Fig.~\ref{fig:twotimes} shows the temporal decay of the persistence probability $\pl(t,t_0)$ in a critical 1+1-dimensional contact process as a function of $t$ for various values of~$t_0$. As can be seen, by increasing $t_0$ the initial transient becomes more pronounced. Moreover, the curves are shifted to the right. However, the asymptotic slope, i.e., the persistence exponent $\tl$, is not affected.

In order to describe this behavior by appropriate scaling laws, the straight-forward method would be to include both parameters $t_0$ and $t$ in the scaling form and to require invariance under scale transformations of the form
\begin{eqnarray}
\label{ScaleTransformation}
&&\Delta 	\to \Lambda \Delta\,, \hspace{14mm}
\rho    \to \Lambda^\beta \rho\,, \hspace{14mm}
L       \to \Lambda^{-\nu_\perp} L\\
&&t_0     \to \Lambda^{-\nu_\parallel}t_0\,, \hspace{10mm}
t       \to \Lambda^{-\nu_\parallel}t\,, \hspace{10mm}
\pl     \to \Lambda^{\kappa} \pl \nonumber , 
\end{eqnarray} 
where $\Lambda>0$ is a scale factor, $\Delta=\lambda-\lambda_c$ is the distance from criticality, $\rho$ denotes the density of active sites, $L$ is the system size, and $\kappa$ is an exponent to be determined below. At criticality in an infinite system, invariance under this set of transformations would lead to the scaling form $\pl(t,t_0) = t^{-\kappa/\nu_\parallel} \, F(t/t_0)$ with an unknown exponent~$\kappa$. This exponent can be determined by keeping the difference $\Delta t= t-t_0$ fixed and taking $t$ to infinity. In this limit the critical system approaches the absorbing state and therefore the probability $\pl(t,t_0)$ to find \textit{no} activity between $t_0$ and $t_0+\Delta t$ tends to $1$. This implies that $\kappa=0$, i.e., the local persistence probability is a quantity that does not carry an intrinsic scaling dimension. Therefore, we arrive at the scaling form
\begin{equation}
\label{CriticalScalingForm}
\pl(t,t_0) = F(t/t_0) \,,
\end{equation}
where $F$ is a scaling function with the asymptotic behavior
\begin{eqnarray}
\label{Asymptotics}
F(z) &\sim z^{-\tl}& \qquad \mbox{ for } z \gg 1, \\
F(z) &\to   1      & \qquad \mbox{ for } z \to 1. \nonumber
\end{eqnarray} 
It is interesting to note that autocorrelation and autoresponse functions in non-stationary critical systems behave similarly. These two-time quantities were studied in recent years in the context of aging~\cite{Hinrichsen07} and exhibit similar scaling laws, of course with different exponents and scaling functions.

In order to verify the scaling form~(\ref{CriticalScalingForm}) and its asymptotic behavior~(\ref{Asymptotics})  we simulated a 1+1-dimensional contact process and measured the probability that a given site is not activated in the time interval between $t_0$ and $t$. The left graph in Fig.~\ref{fig:twotimes} shows curves for fixed $t_0$ as a function of $t$. The graph on the right side demonstrates the data collapse according to the scaling form (\ref{CriticalScalingForm}), plotting $\pl(t,t_0)$ versus the quotient $t/t_0-1$. Although the collapse seems to be convincing, it is not really perfect. The reason will become clear in the following subsection.

\subsection{Graded persistence}
%
In 1996 Ben-Naim {\it et al.}~\cite{BenNaim96} introduced a parametrized generalization of local persistence, namely, the probability $\hat P_n(t)$ that a given site changes its state exactly $n$ times until time $t$. Clearly, for $n=0$ one retrieves the usual definition of local persistence. Studying the voter model the authors showed that this quantity obeys the scaling form
\begin{equation}
\label{BenNaimEtAl}
\hat P_n(t) \sim \frac{1}{\sqrt{t}}\Phi\Bigl(\frac{n}{\sqrt{t}}\Bigr).
\end{equation}
Later, Majumdar and Bray~\cite{Majumdar98b} considered a similar generalization called \textit{persistence with partial survival} $\bar P(p ,t)$, where upon a spin flip a site is decleared as non-persistent with probability $p$. This definition reduces to the original one in the limit $p \to 1$ and is related to the previous generalization by $\bar P(p,t)=\sum_{n=0}^\infty p^n \hat P_n(t)$. 

In addition, Majumdar and Bray argued that there are in principle two different scenarios: If the underlying process is `smooth', meaning that the average number of zero crossings (spin flips) per site $\langle n \rangle$ is finite, one has $\bar P(p,t)\sim t^{-\theta_p}$ with a generalized local persistence exponent $\theta_p$ that varies continuously with $p$. This situation is observed e.g. in the diffusion equation starting with a random intial field. On the other hand, if the underlying process is `non-smooth', meaning that $\langle n \rangle$ is infinite, there is no continuously varying persistence exponent, instead the scaling behavior discussed by Ben-Naim {\it et al.} is recovered. This happens e.g. in the Glauber-Ising model. In the case of directed percolation the average number of spin flips per site is directly linked to the integrated particle density $\langle n \rangle\sim\int_0^\infty {\rm d} t\, \rho(t)$, hence $\langle n \rangle$ is infinite below the upper critical dimension so that the scenario of a `non-smooth' process should apply. 

It is interesting to note that for DP such a generalized persistence probability can also be motivated by field-theoretical arguments as follows. In a continuum formulation, the discrete variables of the lattice $s_i(t)$ are replaced by a coarse-grained density $\rho(\rvec,t)$ of active sites. As suggested by Cardy~\cite{Cardy95}, a suitable definition of the local persistence probability in the continuum limit is given by
\begin{equation}
\label{ContinuousPersistence}
\pl(t,t_0,L,\Delta,\mu) \;=\; \left\langle \exp \left( 
-\frac{1}{\mu} \, \int_{t_0}^{t} \mathrm{d}t' \, \rho(\rvec,t') \right) \right\rangle \,.
\end{equation}
In this expression, which is also needed to describe epidemic process with immunization~\cite{Cardy83,CardyGrassberger85,Janssen85,DammerHinrichsen03,JimenezHinrichsen03,DammerHinrichsen04}, an additional parameter $\mu$ has to be introduced in order to make the argument of the exponential function dimensionless. This parameter plays a similar role as $n$ in Eq.~(\ref{BenNaimEtAl}) and may be interpreted as a typical threshold for the integrated activity below which the site at position $\rvec$ is `declared' as being persistent. 

Invariance of the argument of the exponential function under scale transformation~(\ref{ScaleTransformation}) requires that the parameter $\mu$ scales as
\begin{equation}
\mu \to \Lambda^{\beta-\nu_\parallel} \mu
\end{equation}
so that $\pl$ is a generalized homogeneous function of the form
\begin{equation}
\pl(t,t_0,L,\Delta,\mu) \;=\; \pl(
\Lambda^{-\nu_\parallel}t,\,
\Lambda^{-\nu_\parallel}t_0,\, 
\Lambda^{-\nu_\perp}L,\, 
\Lambda\Delta,\, 
\Lambda^{\beta-\nu_\parallel}\mu
)\,.
\end{equation}
Therefore, a scaling form for the local persistence probability, similar to the one in Eq.~(\ref{BenNaimEtAl}), is given by
\begin{equation}
\label{PostulatedScaling}
\pl(t,t_0,L,\Delta,\mu) \;=\; F\Bigl(
\frac{t}{t_0},\,
\frac{t}{L^z},\,
\frac{t}{\Delta^{-\nu_\parallel}},\,
\frac{t}{\mu^{1/(1-\alpha)}}
\Bigr)\,,
\end{equation}
where $\alpha=\beta/\nu_\parallel$. For $L\to\infty$, $\Delta\to 0$, and $\mu\to 0$ this scaling form reduces to Eq.~(\ref{CriticalScalingForm}), which would predict a perfect data collapse in Fig.~\ref{fig:twotimes}. However, in most numerical simulations, $\mu$ is effectively given by a small but finite number of order $1$. Thus, dropping the last argument in Eq.~(\ref{PostulatedScaling}) will lead to small errors. This is probably the reason why the data collapse shown in Fig.~\ref{fig:twotimes} is quite satisfactory but not perfect.

\begin{figure}
\centering\includegraphics[width=120mm] {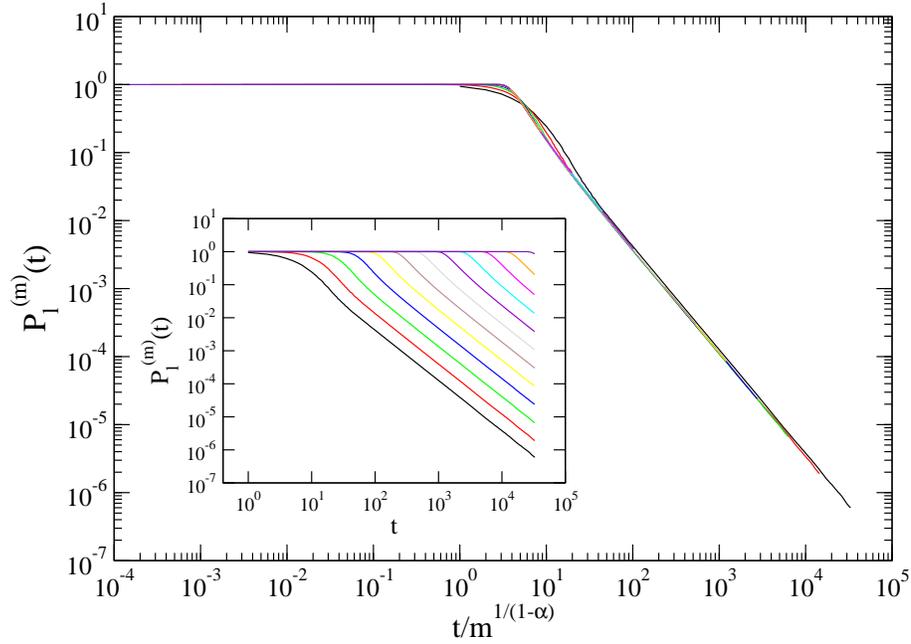}
\caption{\small Graded persistence. The figure shows measurements of the probability $P^{(m)}(t)$ that a given site is activated less than $m$ times at time~$t$. The inset shows the original data while the main graph demonstrates the corresponding data collapse.}
\label{fig:graded}
\end{figure} 

In order to verify this scaling law quantitatively, we consider a \textit{graded persistence probability}
\begin{equation}
\pl^{(m)}(t,t_0) = \left\langle 
\chi\Bigl(m-\sum_{t'=t_0}^{t} s_i(t')\Bigr)
\right\rangle\,,
\end{equation} 
where $s_i(t)=0,1$ is the state of site $i$ and 
\begin{equation}
\chi(z) \;=\; \left\{
\begin{array}{ll}
1 & \mbox{ if } z \geq 0 \\
0 & \mbox{ if } z < 0 
\end{array}
\right. 
\end{equation} 
is a step function. Obviously $\pl^{(m)}(t,t_0)$ is the probability that a given site $i$ is activated \textit{less} than $m$ times during the time interval from $t_0$ until $t$. For $t_0=0$ it is related to the generalization by Ben-Naim {\it et al.} by $\pl^{(m)}(t,0)=\sum_{n=0}^{m-1}\hat P_n(t)$. The reason why we define graded persistence as the probability for having \textit{less} than $m$ activations is that the parameter $m \in \NN$ plays the same role as the constant $\mu$ in Eq.~(\ref{ContinuousPersistence}). 

To verify the postulated scaling behavior in Eq.~(\ref{PostulatedScaling}) we measured $\pl^{(m)}(t,t_0)$ for $t_0=1$ in a critical system for various values of $m$. As demonstrated in Fig.~\ref{fig:graded}, by plotting $\pl^{(m)}(t)$ versus $t/m^{1/(1-\alpha)}$ all curves collapse onto a single one. This data collapse confirms \textit{a posteriori} the field-theoretic definition of local persistence in Eq.~(\ref{ContinuousPersistence}). 

\section{Mapping local persistence to the problem of a permanently active site}
\label{WetSection}

The field-theoretic action describing DP transitions is known to be invariant under the so-called rapidity reversal symmetry~\cite{Grassberger79}, implying that the density of active sites $\rho(t)$ in systems with homogeneous initial conditions and the survival probability $P(t)$ in seed simulations exhibit the same type of asymptotic power-law scaling. In bond directed percolation, this symmetry even holds exactly on the microscopic level since for a given realization of open and closed bonds the probability to find a directed path from site $A$ to site $B$ forward in time and the corresponding probability to find a directed path from $B$ to $A$ \textit{backward} in time are identical. In the following let us consider the implication of the rapidity reversal symmetry to the problem of local persistence in DP. 

\begin{figure}
\begin{flushright}
\includegraphics[width=130mm] {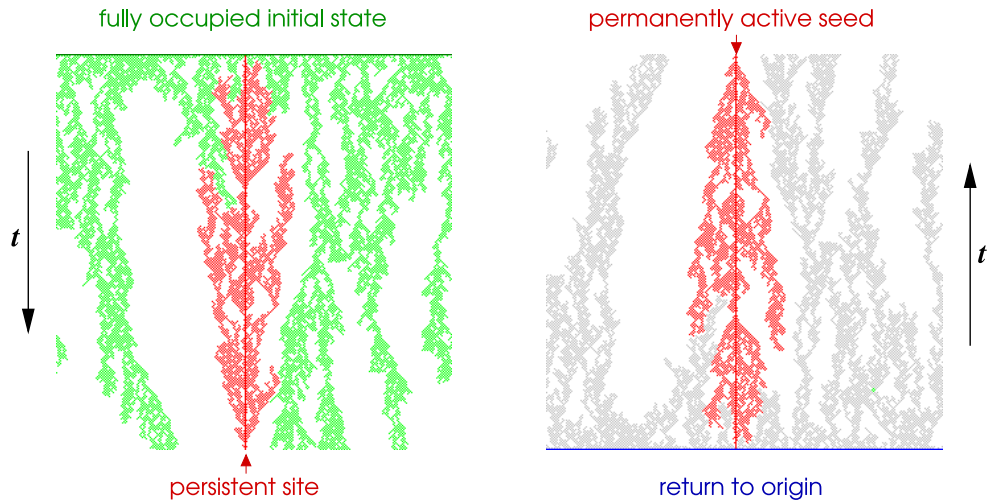}
\end{flushright}\caption{\small Rapidity reversal symmetry: Reversing the time arrow the local persistence probability $\pl(t)$ in bond-DP is exactly equal to the return probability of a cluster generated by a permanently active seed (see text).}
\label{fig:activeseed}
\end{figure} 

\subsection{Relation to a DP process with a permanently active source}
%
As illustrated in left panel of Fig.~\ref{fig:activeseed}, a site is locally persistent (vertical red line) if it is not reached by any directed path from the initial state (upper horizontal green line) until time $t$. This means that the cluster generated by the initial state (green pixels) and the set of all directed paths leading to the persistent site (red pixels) are disjoint. 

The right panel of Fig.~\ref{fig:activeseed} shows exactly the same realization of open and closed bonds turned upside-down which can be interpreted as a bond-DP process running backward in time. As can be seen, the red cluster may now be interpreted as being generated by a permanently active site in the middle. Obviously, the local persistence probability is exactly equal to the probability $R(t)$ that this cluster does not reach any site at the final time (horizontal blue line). In other words, $\pl(t)$ coincides with the \textit{return probability} $R(t)$ of a DP cluster generated by a permanently active site in an otherwise empty system to its initial condition (i.e. an empty lattice but the single permanently active site) after $t$ time steps.

In other realizations of DP, e.g. in the contact process, the time reversal symmetry does not hold exactly on the microscopic level, instead it emerges as an asymptotic symmetry on large scales, just in the same sense as a random walk on a square lattice becomes rotationally invariant on large scales. Therefore, apart from a non-universal proportionality constant, the relation between local persistence and the return probability of clusters generated by a permanently active seed is expected to be valid in all models belonging to the DP universality class \cite{Hinrichsen98a}.

\subsection{One-sided persistence in one spatial dimensions}
\begin{figure}
\begin{flushright}
\includegraphics[width=130mm]{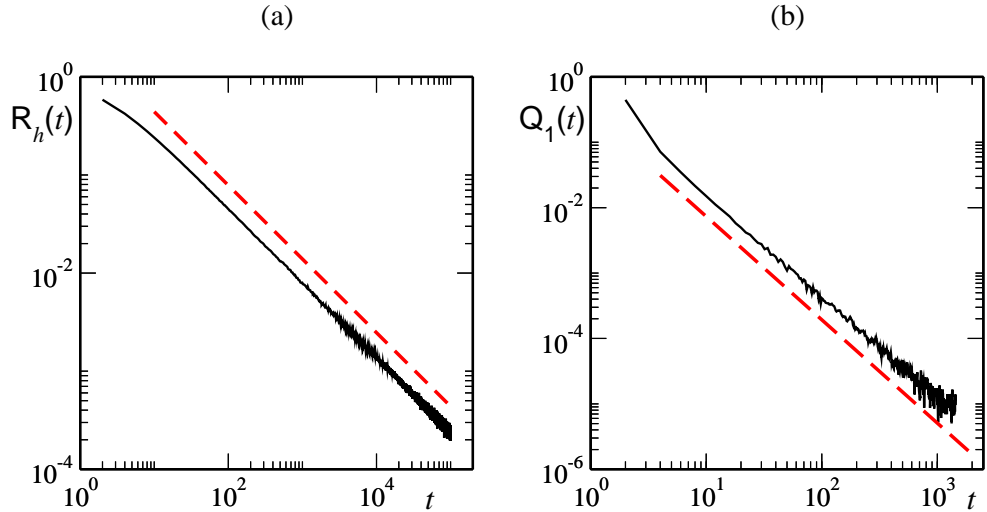}
\vspace{-2mm}
\end{flushright}\caption{\small (a) One-sided return probability $R_{h}(t)$ of clusters generated by a permanently active seed for critical bond directed percolation confined in the semiplane of a (1+1)-dimensional tilted square lattice (see text). The dashed red line marks the power low decay $t^{-\tl/2}$ with $\tl = 1.517$. Data has been averaged over $4\cdot10^5$ independent realizations. (b) First-return probability distribution $Q_1(t)$ of clusters generated by a permanently active seed for critical bond directed percolation on a (2+1)-dimensional body-centered cubic lattice. The red dashed line indicates the lower bound for $\tl$ in two spatial dimensions reported in Table~\ref{Table:Exponents}. Data has been averaged over $2.5\cdot10^5$ independent realizations.}
\label{FigWet}
\end{figure}

%
In one spatial dimension, the permanently active seed splits the lattice in two disjoint uncorrelated domains, so that the return probability to the initial condition in the entire space is equal to the square of the one-sided return probability when only a single semispace is considered, i.e.  $R_{h} (t) \sim t^{-\tl/2}$. In principle, this would allow for a better precision in numerical simulations, since a slower decay would lead to a better statistics for large times. A measurement of $R_{h}(t)$ at the critical point of bond directed percolation is shown in Fig.~\ref{FigWet}a, confirming the halved value of the persistence exponent. In higher spatial dimensions, the permanently active seed no longer splits the space into disjoint domains, and so one has to deal with the return probability to the initial condition in the full space. 

\subsection{Return vs. first-return probability}
%
In order to verify the identity between local persistence probability and the probability to return to the initial condition of clusters generated by a permanently active seed in higher dimensions, we chose to measure the \textit{first}-return probability distribution $Q_1$ which is defined as follows: Consider an active seed located at the origin of a lattice in $d$ spatial dimensions. Let $Q_1(t)$ be the probability for the cluster generated from the active seed located to return to its initial condition at time $t$ {\it for the first time}, with $Q_1(t)=0$ for $t<=0$. Then it follows that the probability $Q_2(t)$ of the second return to the initial state, the last one at time $t$, is given by the discrete convolution
\begin{equation}
Q_2(t) = \sum_{n=1}^\infty Q_1(n) Q_1(t-n) = (Q_1 * Q_1)(t)\,
\end{equation}
since successive returns to the initial state are uncorrelated. This can be extended recursively to the probability of $k$ returns (with $k \geq 2$), i.e. $Q_k(t) =  (Q_{k-1} * Q_1)(t)$. Suppose that $Q_1(t) \sim t^{-\gamma}$ with $\gamma>1$. By the convolution theorem for the Laplace transform one has the asymptotic $Q_k(t) \sim t^{(1-\gamma) k -1}$. It is now easy to verify that the unrestricted return probability $R(t)$ can be expressed as a sum of the $k$ first-return probability distributions,
\begin{equation}
R(t) = \sum_{k=1}^t Q_k(t) 
\label{sumk}
\end{equation}
which scales asymptotically as $t^{-\gamma}$. Therefore, to leading order, from the rapidity reversal argument one simply expects
\begin{equation}
\gamma = \tl \,.
\end{equation}
In Fig.~\ref{FigWet}b the decay for the first-return probability $Q_1(t)$ at the critical point of two spatial dimension bond directed percolation is compared with the lower bound reported in Table~\ref{Table:Exponents}.

\section{Local persistence in seed simulations}
%
Up to now we considered only a fully occupied lattice $\rho_0 = 1$ as initial configuration. More generally one could start with a finite density $0<\rho_0<1$ of active sites. In \cite{Hinrichsen98a} it is shown that in 1+1-dimensional DP, after an initial transient, the local persistence probability follows the power law $\pl(t)\sim t^{-\tl}$ with an exponent $\tl$ which is independent of $\rho_0$. In the limit $\rho_0 \rightarrow 0$ the transient regime takes an infinite amount of time, meaning that asymptotic power law is no longer accessible.

\begin{figure}
\begin{flushright}
\includegraphics[width=130mm] {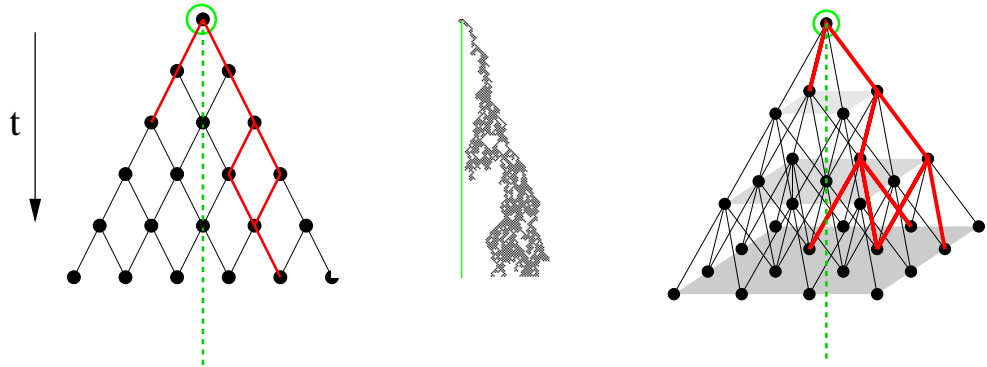}
\vspace{-5mm}
\end{flushright}\caption{\small Local persistence in seed simulations. Left: Lattice configuration in (1+1) dimensions. Each lattice site has two nearest neighbors, which could be activated in the next time step, following the red path. A first-return to the origin is only possible at even times, starting with one single seed at the origin at time $t=0$ (green). Middle: Snapshot of the evolution of a surviving percolation cluster in one spatial dimension. Only the right half of the cluster is active. Right: Lattice configuration in (2+1) dimensions. Again, possible paths (red) and the seed at the origin (green) are shown.}
\label{fig:seed}
\end{figure}

Another well-known simulation technique starts with a single `seed' at the origin of the lattice, which generates a certain cluster of active sites (see Fig.~\ref{fig:seed}). This simulation technique was first introduced in 1979 by Grassberger and de la Torre~\cite{Grassberger79}. Many of these clusters will die out soon, but some of them will survive for long time. The advantage of this approach is that the number of particles at each time step is finite. Thus, instead of storing the whole lattice configuration one only has to keep track of the coordinates of the active sites in a dynamically generated list. Apart from the range of integer numbers, this approach imposes no boundary effects, thus one effectively simulates an infinite system. 

It is interesting to investigate how the local persistence probability $P_l^{\scriptscriptstyle \rm seed}(t)$ behaves in simulations starting with a single seed. In this case local persistence may be defined as the probability that the origin, from where the cluster is generated, is not activated until time $t$. However, since a finite fraction of clusters dies before they have returned to the origin, this quantity, when averaged over \textit{all} runs, would tend to a constant for $t \rightarrow \infty$. Therefore, local persistence should be defined as an average over the \textit{surviving} clusters. The question is, whether 
\begin{itemize}
\item[(i)] $P_l^{\scriptscriptstyle \rm seed}(t)$ follows a power-law behavior as $P_l(t)$ does in a system with homogeneous initial conditions, and 
\item[(ii)] if so, whether and how the corresponding exponent $\tl^{\scriptscriptstyle \rm seed}$ is related to $\tl$.
\end{itemize}
Equivalently one can measure the probability distribution\footnote{In \cite{Albano01}, which deals with homogeneous initial conditions, the corresponding quantity $D_1(t)$ is called the \textit{persistence probability distribution}, which gives the probability for a persistent site to become active in the time interval between $t$ and $t + dt$.} $D_1^{\scriptscriptstyle \rm seed}(t)$ for the first return of activity to the origin, which essentially is the time derivative of $P_l^{\scriptscriptstyle \rm seed}(t)$,
\begin{equation}
\frac{d}{dt}P_l^{\scriptscriptstyle \rm seed}(t)  = - D_1^{\scriptscriptstyle \rm seed}(t) .
\end{equation}
Note that $D_1^{\scriptscriptstyle \rm seed}(t)$ differs from the first return probability distribution $Q_1(t)$ considered in the Sect.~\ref{WetSection} in so far as the latter refers to the return to the original initial condition when the seed is kept permanently active in time, while the former samples first-reactivation times of a normal seed.

Assuming a power-law behavior, we expect a different value for the local persistence exponent $\tl^{\scriptscriptstyle \rm seed}$for the following reason: The initial condition of a single seed can be interpreted as a homogeneous initial state in the limit $\rho_0 \rightarrow 0$, with an important difference: We only investigate the first-return probability of a single seed at the origin, which is a non-homogeneous problem. In a sense, we exclude all lattice sites except the origin from the calculation of the first-return probability. Consequently, $D_1^{\scriptscriptstyle \rm seed}(t) < D_1(t)$ (for all $t$), which immediately leads to $\tl^{\scriptscriptstyle \rm seed} \geq \tl$. In addition the clusters have a finite lifetime so that $D_1^{\scriptscriptstyle \rm seed}(t)$ is again expected to decline faster than $D_1(t)$. Therefore it is natural to expect $\tl^{\scriptscriptstyle \rm seed}$ to be greater than $\tl$. Moreover, persistence in seed simulations is not defined in the limit of infinitely many dimensions since then the generated cluster never returns back to the origin.

\begin{figure}
\begin{center}
\includegraphics[width=90mm] {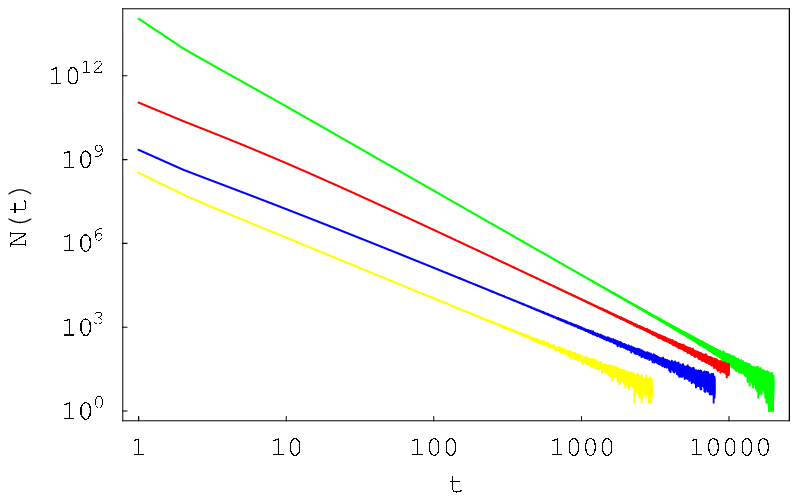}
\hspace{-26mm}
\includegraphics[width=90mm] {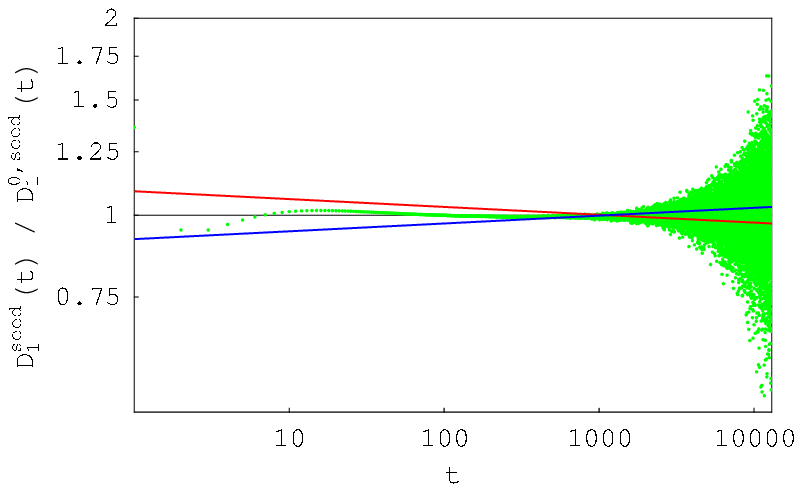}
\vspace{-10mm}
\end{center}
\caption{Local persistence in seed simulations. Left: Time distribution of the frequency $N(t)$ of the first-return in 1 (green), 2 (red), 3 (blue) and 4 (yellow) spatial dimensions. For better visualization neighboring data points are joined. Right: One-dimensional data $D_1^{\scriptscriptstyle \rm seed}(t)$ divided by the fitted power law $D_1^{\scriptscriptstyle \rm 0,seed}(t) \propto t^{-\tl^{\scriptscriptstyle \rm seed}}$ plotted against time $t$ (green). The red and the blue line indicate the error margins for the estimate of $\tl^{\scriptscriptstyle \rm seed}$.}
\label{fig:seeddata}
\end{figure} 

In order to verify this conjecture we simulated the bond-DP process on a simple cubic lattice in $d = 1$ to $4$ spatial dimensions. In this model, each lattice site has $2d$ nearest neighbors. We start with a single seed at the origin. Notice that because of the lattice structure, a first-return to the origin can only occur at even time steps (see Fig.~\ref{fig:seed}). Many of the evolving percolation clusters die out before a first-return to the origin occurs (approximately $49\%$ in one, $56\%$ in two, $74\%$ in three and $83\%$ in four spatial dimensions). We simulated about $10^{10}$ independent runs in dimensions two to four and, in order to increase the precision of $\tl^{seed, 1D}$, about $10^{14}$ independent runs in the one-dimensional case. 

The results are displayed in Fig.~\ref{fig:seeddata}. In the one-dimensional case the evaluation of the data yields a clean power-law for the first-return-probability distribution $D_1^{\scriptscriptstyle \rm seed}(t)$ with exponent $\tl^{seed, 1D} + 1 = 3.014 \pm 0.012$. Although the integer value 3 lies only slightly outside of the error margins and we therefore cannot exclude this value for sure, we see no reason for $\tl^{seed, 1D}$ to be integer. The higher-dimensional cases also appear to follow a power-law behavior. However, a more detailed analysis by plotting the data divided by the fitted power-law shows systematic deviations. If we anticipate a power-law, limits for the corresponding exponents are given by $\tl^{\scriptscriptstyle \rm seed, 2D} + 1 > 2.5$, $\tl^{\scriptscriptstyle \rm seed, 3D} + 1 > 2.2$ and $\tl^{\scriptscriptstyle \rm seed, 4D} + 1 \approx 2$. The question whether $\tl^{\scriptscriptstyle \rm seed}$ is related to $\tl$ is still open.

%
\section{Temporal vs. spatial persistence}
%

So far we defined local persistence as a time-dependent probability $\pl(t)$ that a given site remains inactive from time $t_0=1$ until time $t$. For homogeneous initial conditions $\pl(t)$ decays as $t^{-\tl}$, where $\tl$ is the persistence exponents. Since space and time play a similar (although not a symmetric) role in DP, the question arises whether it is possible to consider a \textit{spatial} persistence probability $\tilde{P}(r)$ with analogous properties. Obviously, such a spatial persistence probability should be defined as the probability that a \textit{line} of sites extending from the origin to site $r$ next to an active seed is found to be inactive. The problem of spatial persistence was first introduced and analyzed by Majumdar and Bray in the context of fluctuating interfaces~\cite{MajumdarBray01}.

\begin{figure}
\begin{center}
\includegraphics[width=105mm] {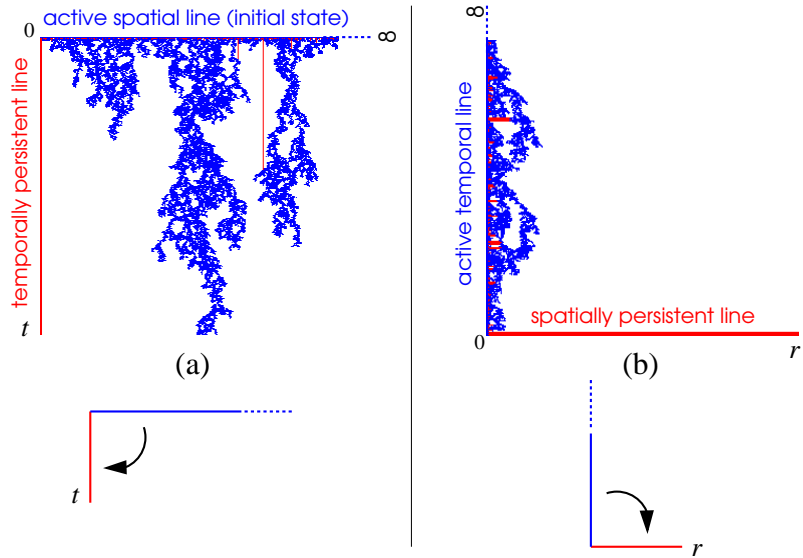}
\end{center}
\vspace{-5mm}
\caption{Temporal vs. spatial persistence in 1+1 dimensions (see text).
On both sides time runs in downward direction.}
\label{fig:spatialdemo}
\end{figure} 

As shown in Fig.~\ref{fig:spatialdemo} spatial persistence in DP is most easily introduced in 1+1 dimensions. Part~(a) of the figure shows the one-sided version of ordinary temporal persistence: Starting with an initial state where all sites to the right of the origin are occupied, the one-sided persistence probability $P_h(t)$ is the probability that no path from the horizontal line of initially occupied sites reaches the vertical line at the origin extending up to time $t$. This probability is known to decay as $P_h(t) \sim t^{-\tl/2}$, where $\tl$ is the local persistence exponent.

Part (b) shows the same arrangement rotated by $90^\circ$: Here we ask for the probability $\tilde{P}_h(r)$ that there is no path from a vertical line at the origin $-\infty < t < 0$ to the finite horizontal line extending from the origin to site $r$. More specifically, we keep the site at the origin active and let the system evolve. Once it reaches a stationary state we ask for the probability that the sites $1,2,\ldots,r$ are inactive. Obviously, this quantity is a spatial analogue of temporal persistence.

\begin{figure}
\begin{center}
\includegraphics[width=110mm] {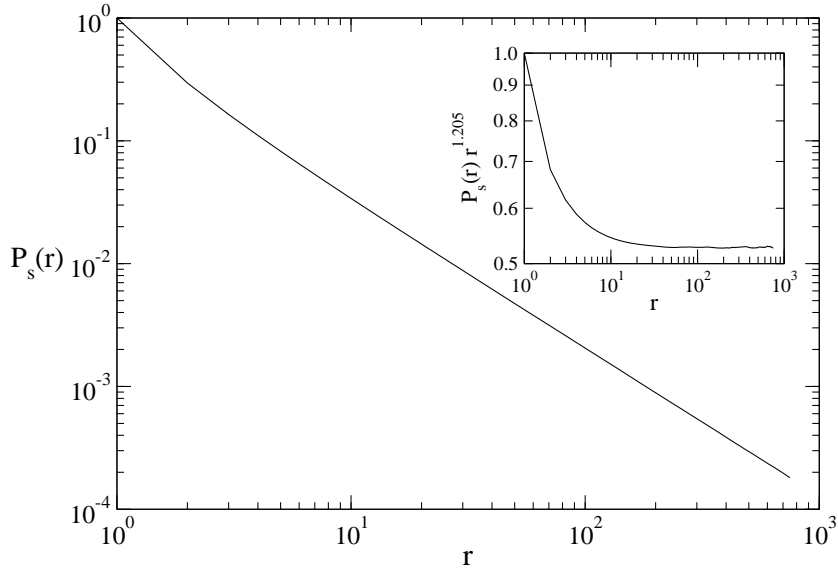}
\end{center}
\caption{Monte Carlo simulation of spatial persistence in a 1+1-dimensional contact process at criticality.}
\label{fig:spatialpers}
\end{figure}

Fig.~\ref{fig:spatialpers} shows a numerical measurement of $\tilde{P}_h(r)$ in the stationary state of a critical contact process with an active seed at the origin. We observe that the one-sided spatial persistence probability decays algebraically as
\begin{equation}
\tilde{P}_h(r) \sim r^{-\kappa/2}\,, \qquad \kappa/2 = 1.205(8)
\end{equation}
This estimate is consistent with the scaling relation
\begin{equation}
\kappa = \tl\,z\,,
\end{equation}
where $z=\nu_\parallel/\nu_\perp$ is the dynamical exponent of DP. Clearly, the two-sided spatial persistence probability that $r$ sites on both sides of the active seed are inactive will scale as $\tilde{P}(r) \sim r^{-\kappa}$.

In dimensions $d>1$ spatial persistence may be defined as the probability in a stationary DP process with an active seed at the origin that all sites in a $d$-dimensional sphere of radius $r$ surrounding the origin are inactive.

%
\section{Concluding remarks}
%
In this work we have analyzed in detail different aspects of local persistence in DP processes. Concerning persistence in $d>d_c=4$ spatial dimensions, a simple mean field approach shows that the local persistence exponent $\tl$ is a non-universal quantity related to the amplitude of the density field, which generally depends on both dimensionality and model details. A similar non-universal behavior above the upper critical dimension was previously observed in other reaction diffusion models as well~\cite{Cardy95}. 

Below the upper critical dimension, we have performed high precision numerical simulations in order to improve current estimates of $\tl$ in one spatial dimension. On this basis we can safely exclude the conjecture $\tl=3/2$, i.e., the local persistence exponent -- like other DP critical exponents -- is seemingly not given by a simple rational value.

The simulations in $d=2,3,4$ spatial dimensions are less conclusive, apparently due to huge transient effects, which make it impossible to estimate the value of $\tl$ by numerical simulations. However, specifying lower bounds we can at least rule out a possible {\it superuniversality} between $d=1$ and $d=2$ mentioned in \cite{Albano01}. Although we have not been able to determine $\tl$ in $d < 4$, or even to prove the existence of an asymptotic power law behavior analytically, we have discussed various concepts -- scaling relations, graded persistence, mapping to active seed problems and spatial persistence -- which we hope could prove useful for further theoretical and numerical advances.

Finally, let us add a comment on our results concerning local persistence in seed simulations. Numerical results exhibit a clean power-law in the one-dimensional case, whereas in higher dimensions we observe systematic deviations from a power-law behavior within the numerically accessible range of time. It is not clear whether these deviations just resemble a longer-lasting transient period. They could also be related to a qualitative difference between the one-dimensional and the higher-dimensional cases, or more precisely to a kind of special screening effect in one dimension: In one dimensional seed simulations the persistence of the lattice site at the origin is not influenced by remote branches of the emerging cluster, because these are shielded by the branches next to the origin (left and right), as long as those have not died out. If a remote branch approaches the origin, it will merge with the inner branches without changing their dynamics (this is particularly easy to see in 1+1-dimensional directed bond percolation). In other words, a return of the seed to the origin can only be accomplished from one of the two branches next to the origin. In higher dimensions, there is no such shielding effect. Take for example the (2+1)-dimensional case: The initial seed develops into a cone-like cluster, and the origin could be reached from any direction in the plane of the lattice, not only from the nearest branches. So, with the breakdown of the shielding effect, a return to the origin is more likely, which leads to a stronger, perhaps even non-algebraic decline of local persistence.

\ack
We wish to thank I. Dornic for many friendly and stimulating discussions.

\vspace{2mm}
\noindent{\bf References}\\

\end{document}